\documentstyle[aps,psfig,epsf]{revtex}

\input epsf

\tightenlines

\begin{document}

\title{\Large{Quenching of Hadron Spectra due to the Collisional Energy Loss 
of Partons in the Quark-Gluon Plasma}}

\author{Munshi G. Mustafa$^1$ and 
Markus H. Thoma$^2$ }

\address{$^1$Theory Group, Saha Institute of Nuclear Physics, 1/AF Bidhan 
Nagar, Kolkata 700 064, India}

\address{$^2$ Centre for Interdisciplinary Plasma Science, 
Max-Planck-Institut f\"ur extraterrestrische Physik,
P.O. Box 1312, 85741 Garching, Germany}

\maketitle

\vspace{0.2in}

\begin{abstract}
We estimate the energy loss distribution and investigate the quenching of 
hadron 
spectra in ultrarelativistic heavy-ion collisions due to the collisional energy
loss of energetic partons from hard parton collisions in the initial stage.

\end{abstract}

\vspace{0.2in}


\vspace{0.2in}


\vspace{0.2in}

In the initial stage of ultrarelativistic heavy-ion collisions energetic 
partons are produced from hard collisions between the partons of the nuclei.
Receiving a large transverse momentum, 
these partons will propagate through the 
fireball which might consist of a quark-gluon phase for a transitional period
of a few fm/c. These high-energy partons will manifest themselves as jets
leaving the fireball. Owing to the interaction of the hard partons with
the fireball these partons will lose energy. Hence jet quenching will result.
The amount of quenching might depend on the state of matter of the fireball, 
i.e., quark-gluon plasma (QGP) or a hot hadron gas, respectively. Therefore
jet quenching has been proposed as a possible signature for the QGP formation
\cite{Pluemer}. 

In order to use jet quenching as a signature for the quark-gluon plasma,
the energy loss of hard partons in the QGP has to be determined. First
the energy loss due to collisional scattering was estimated 
\cite{Bjorken,svet,Thoma1,Mrowczynski,Koike}.  
Using the Hard-Thermal-Loop
(HTL) resummed perturbative QCD at finite temperature \cite{BP}, the 
collisional energy loss of a heavy quark 
could be derived in a systematic way \cite{Braaten1,Braaten2,Vija,Romatschke}.
From these results also an estimate for the 
collisional energy loss of energetic gluons and light quarks
could be derived \cite{Thoma2}, which
was rederived later using the Leontovich relation \cite{Thoma3,Thoma4}.  
Later also the energy loss due to multiple gluon radiation (bremsstrahlung)
was estimated and shown to be the dominant process. For a review 
on the radiative energy loss see Ref. \cite{BSZ}. Recently, it has also been 
shown~\cite{WW} that for a moderate value of the parton energy there is a
net reduction in the parton energy loss induced by multiple scattering due
to a partial cancellation between stimulated emission and thermal
absorption. This can cause a reduction of the quenching factor due to  
radiative processes.

Unfortunately, jets, requiring very large initial parton energies, 
are rare events, which are 
difficult to observe. However, quenching of hard partons 
will also affect hadron spectra at high transverse momenta $p_\bot$. Indeed,
first results from RHIC have indicated a suppression of high-$p_\bot$ spectra
\cite{RHIC}. The consequences of jet quenching on hadron spectra have been 
calculated, for example, in Refs.\cite{Baier,muel}. Here only the 
radiative energy loss has been taken into account. The purpose of the present 
paper is to estimate the quenching of hadron spectra due to the collisional
energy loss of partons in the QGP. As we will see, this contribution can
be of the same order as the radiative quenching.

We will follow the investigations by Baier et al. \cite{Baier} and 
M\"uller \cite{muel}, using the collisional instead of the radiative
parton energy loss. Following Ref.\cite{Baier} the $p_\bot$ distribution 
is given by the convolution of the transverse momentum distribution in 
elementary hadron-hadron collisions, evaluated at a shifted value 
$p_\bot+\epsilon$, with the probability distribution, $D(\epsilon )$,
in the energy $\epsilon$, lost by the partons to the medium by collisions,
as
\begin{eqnarray}
\frac{d\sigma^{\rm{med}}}{d^2p_\bot} &=& \int d\epsilon \, D(\epsilon) \,
\frac{d\sigma^{\rm{vac}}(p_\bot+\epsilon)}{d^2p_\bot}  
= \int d\epsilon \, D(\epsilon) \,
\frac{d\sigma^{\rm{vac}}}{d^2p_\bot}
+\int d\epsilon \, D(\epsilon) \, \epsilon \,
\frac{d}{dp_\bot} \frac{d\sigma^{\rm{vac}}}{d^2p_\bot}
+\cdots \cdots \nonumber \\
&=& \frac{d\sigma^{\rm{vac}}}{d^2p_\bot} + \Delta E \cdot 
\frac{d}{dp_\bot} \frac{d\sigma^{\rm{vac}}}{d^2p_\bot}
=  \frac{d\sigma^{\rm{vac}}(p_\bot+\Delta E)}{d^2p_\bot} 
= Q(p_\bot) \frac{d\sigma^{\rm{vac}}(p_\bot)}{d^2p_\bot}.
\label{rate}
\end{eqnarray}
Here $Q(p_\bot)$ is suppression factor due to the medium and 
the total energy loss by partons in the medium is
\begin{equation}
\Delta E = \int \epsilon \, D(\epsilon) \, d\epsilon \, . \label{eloss}
\end{equation}
We need to calculate the probability distribution, $D(\epsilon)$, that a 
parton loses the energy, $\epsilon$, due to the elastic collisions 
in the medium. This requires the evolution of the energy distribution of 
a particle undergoing Brownian motion. The operative equation for the 
Brownian motion of a test particle can be obtained from the Boltzmann 
equation, whose covariant form can be written
as 
\begin{equation}
p^\mu\partial_\mu D(x,p) = C\{ D \} \, , \label{boltz}
\end{equation}
where $p$ is the momentum of the test particle, $C\{ D\}$ is the collision
term and $D(x,p)$ is the distribution due to the motion of the particle. 
If we assume a uniform plasma, the Boltzmann equation becomes
\begin{equation}
\frac{\partial D}{\partial t} = \frac{C\{ D  \}}{E}=
\left ( \frac{\partial D}{\partial t}\right )_{\rm{coll}} \, \, .
\label{unif}
\end{equation}
We intend to consider only the elastic collisions of the test parton with
other partons in the plasma. The rate of collisions $w(p,k)$ is given by
\begin{equation}
w(p,k)= \sum_{j=q, {\bar q}, g} w^j(p,k) \, , \label{coll}
\end{equation} 
where $w^j$ represents the collision rate of a test parton $i$ with other 
partons, $j$, in the plasma. The expression for $w^j$ can be written as
\begin{equation}
w^j(p,k) = \gamma_j \int \ \frac{d^3q}{(2\pi)^3} f_j(q) v_{\rm{rel}} 
\sigma^j \, ,  \label{indiv}
\end{equation}
where $\gamma_j$ is the degeneracy factor, $v_{\rm{rel}}$ is the relative
velocity between the test particle and other participating partons $j$ from
the background, and
$\sigma^j$ is the associated cross section. Due to this scattering the 
momentum of the test particle changes from $p$ to $p-k$. Then the collision
term on the right-hand side of (\ref{unif}) can be written as
\begin{equation}
\left (\frac{\partial D}{\partial t}\right )_{\rm{coll}} = \int \ d^3k \, 
\left [ w(p+k,k)D(p+k) -w(p,k)D(p)\right ] \, \, . \label{transi}
\end{equation}
where the collision term has two contributions. The first one is the
gain term where 
the transition rate $w(p+k,k)$ represents the rate that a particle with
momentum ${\vec p} +\vec k$ loses momentum $\vec k$ due to the reaction with
the medium. The second term represents the loss due to the scattering of
a particle with momentum $\vec p$.

Now under the Landau approximation, i.e., most of the quark and gluon 
scattering is soft which implies that the function $w(p,p^\prime)$ is 
sharply peaked at $p\approx p^\prime$, one can expand the first term on
the right-hand side of (\ref{transi}) by a Taylor series as
\begin{equation}
w(p+k,k)D(p+k) \approx w(p,k)D(p)+ {\mathbf k}\cdot \frac{\partial}
{\partial {\mathbf p}}(wD)+ \frac{1}{2}k_ik_j \frac{\partial^2}{\partial
p_i \partial p_j} (wD)+ \cdots \cdots \, . \label{taylor}
\end{equation}
Combining (\ref{unif}),(\ref{transi}) and (\ref{taylor}), 
the Fokker-Planck equation 
\begin{equation}
\frac{\partial D}{\partial t} = \int d^3k \left [ 
{\mathbf k}\cdot \frac{\partial} {\partial {\mathbf p}}+ 
\frac{1}{2}k_ik_j \frac{\partial^2}{\partial p_i \partial p_j} \right ] (wD)
\label{fp}
\end{equation}
is obtained, which describes the equation of motion for the distribution 
function of fluctuating macroscopic variables.
We consider, for simplicity, the one dimensional problem,  for which
(\ref{fp}) can be written as 
\begin{equation}
\frac{\partial D}{\partial t} = \frac{\partial}{\partial p} 
\left [ {\cal T}_1(p) D \right ] + \frac{\partial^2}{\partial p^2}
\left [ {\cal T}_2(p) D \right ]. \label{fp1}
\end{equation}
Here the transport coefficients are given as
\begin{eqnarray}
{\cal T}_1(p) &=& \int d^3k \, w(p,k)\, k = \frac {\langle \delta p \rangle}
{\delta t} = \langle F \rangle \, , \nonumber \\
{\cal T}_2(p) &=&\frac{1}{2} \int d^3k\, w(p,k)\, k^2 = 
\frac {\langle (\delta p )^2\rangle}
{\delta t} \, \, . \label{coeff}
\end{eqnarray} 
Now the work done by the drag force, ${\cal T}_1$, acting on a test particle is
\begin{equation}
-dE = \langle F\rangle \cdot dL = {\cal T}_1(p)\cdot dL \, \, ,\label{work}
\end{equation}
which can be related to the energy loss~\cite{Braaten2,Thoma2} of a particle as
\begin{equation}
-\frac{dE}{dL} = {\cal T}_1(p) \approx p\, {\cal A}  \, \, , \label{celoss}
\end{equation}
where ${\cal A}$ is the drag coefficient, which is
almost independent of momentum $p$~\cite{svet,munshi}. 
The drag coefficient is a very important quantity containing the
dynamics of elastic collisions. In principle, ${\cal A}(p,t)$, may be
determined from the kinetic theory formulation of QCD through the
application of the fluctuation-dissipation theorem~\cite{bales}, but 
that is indeed a difficult problem. As discussed in 
Refs.~\cite{svet,munshi,jane},
the drag coefficient is expected to be largely determined by the 
properties of the bath and not so much by the nature of the test
particle. Then one can use the average of the drag coefficient given as
\begin{equation}
\langle {\cal A}(p,t) \rangle \, = \, {\cal A}(t) \, = \, 
\left \langle -\frac{1}{p}\, \, \frac{dE}{dL}
 \right \rangle 
\, \, . \label{average}
\end{equation}

The quantity ${\cal T}_2$ can be related
to the diffusion coefficient in the following way: 
\begin{equation}
{\cal T}_2(p) = \frac {\langle (\delta p )^2\rangle}
{\delta t}= p{\cal A} p \approx {\cal A} T^2 = {\cal D}_F \, \, , \label{diff}
\end{equation}
where we have approximated $p$ by the temperature $T$ of the system and the
drag by using the Einstein relation, ${\cal T}_1 T \approx {\cal D}_F$,
assuming that the coupling between the Brownian particle and the bath 
is weak~\cite{bales}. 

Combining (\ref{fp1}), (\ref{celoss}) and (\ref{diff}), we find
\begin{equation}
\frac{\partial D}{\partial t} = {\cal A} \frac{\partial }{\partial p} (p D)
+ {\cal D}_F\frac{\partial^2 D}{\partial p^2} \, \, , \label{landau} 
\end{equation}
which describes the evolution of the momentum distribution of a test particle
undergoing Brownian motion. 

Next we proceed with solving the above equation with the boundary condition
\begin{equation}
D(p,t)\, \, \, \stackrel{t\rightarrow t_0}{\longrightarrow}
\, \, \delta(p-p_0)\, \, . \label{boundc}
\end{equation}

The solution of (\ref{landau}) can be found by making a
Fourier transform of $D(p,t)$,
\begin{equation}
D(p,t)=\frac{1}{2\pi}\int_{-\infty}^{+\infty} \, {\tilde D}(x,t) \, 
\, e^{ipx}dx 
\, \, , \label{fourr}
\end{equation}
where the inverse transform is
\begin{equation}
{\tilde D}(x,t)=\int_{-\infty}^{+\infty} \,  D(p,t) \, \, e^{-ipx}dp 
\, \, . \label{invfourr}
\end{equation}
Under the Fourier transform the corresponding initial condition follows
from (\ref{boundc}) and (\ref{invfourr}) as
\begin{equation}
{\tilde D} (x_0,t=t_0)=e^{-ip_0 x_0} \, \, \label{bcfourr}
\end{equation}
where $x=x_0$ at $t=t_0$ is assumed. 

Replacing 
$p\rightarrow i\frac{\partial}{\partial x}$ and  
$\frac{\partial}{\partial p}\rightarrow ix$, the Fourier transform of
(\ref{landau}) becomes
\begin{equation}
\frac{\partial {\tilde D}}{\partial t} + {\cal A}x \frac{\partial {\tilde D}}
{\partial x} 
= -{\cal D}_F \, x^2 {\tilde D} \, . \label{landau4} 
\end{equation}
This is a first order partial differential equation which may be solved by
the method of characteristics~\cite{william}. The characteristic
equation corresponding to (\ref{landau4}) reads
\begin{equation}
\frac{\partial t}{1} \, = \, \frac{\partial x}{{\cal A} x}\, =\,
 -\frac{\partial {\tilde D}} {{\cal D}_Fx^2{\tilde D}} \, \, \, \, 
. \label{charac}
\end{equation}

Along with the boundary condition in (\ref{bcfourr}) and ${\cal A}(t_0)\, 
= \, {\cal D}_F(t_0) \, = \, 0$, the solution of (\ref{landau4}) can
be obtained as
\begin{equation}
{\tilde D}(x,t) \, =\, \exp\left [ -i\,p_0 \, x (t)\,
 e^{-\int^t{\cal A}(t') dt'}
\right ] \, \exp \left [- \int^t {\cal D}_F(t') 
\, x^2(t')\, dt' \right ] \, \, , \label{sol2}
\end{equation} 
Combining (\ref{fourr}) and (\ref{sol2}) yields
\begin{equation}
D(p,t) = \frac{1}{2\pi} \int_{-\infty}^\infty \, 
\exp\left [ -i\,p_0 \, x (t)\, e^{-\int^t{\cal A}(t') dt'} \right ] \, 
\exp \left [- \int^t {\cal D}_F(t')
\, x^2(t')\, dt' \right ] \, e^{ipx(t)}\, dx \, . \label{solp}
\end{equation}
It is convenient to integrate over $x_0$ instead of $x$ and  
to substitute the solution of the first pair of (\ref{charac}) into the above
equation, leading to
\begin{eqnarray}
D(p,t) &=& \frac{\exp \left (
{\, \int^t{\cal A}(t')\, dt'}\right )}{2\pi} \int_{-\infty}^{+\infty} \, 
\exp\left [ -i\,p_0 \, x_0 \, + \, i\, p x_0\,  e^{\int^t{\cal A}(t') dt'}
\right. \nonumber \\  && \left. \,  -\, x_0^2 
\left \{ \int^t {\cal D}_F(t') \,
\left ( e^{\, 2\int^{t'}{\cal A}(t'')
\, dt'' } \right ) \, dt' \right \} \right ] \,  dx_0 \, . 
\label{sols}
\end{eqnarray}
Using the standard form
\begin{eqnarray}
f(z)\, &=& \, \frac{1}{2\pi} \int_{-\infty}^{+\infty} \, \exp\left[-iuz\, + \,
iuK_1 \, - \, \frac{1}{2} u^2 K_2 \right ]\, du \nonumber \\
&=& \, \frac{1}{\sqrt{2\pi K_2}} \, \exp\left [ -\frac{1}{2} (z-K_1)^2/K_2 
\right ] \, \, , \label{standint}
\end{eqnarray} 
one can easily perform the integration in (\ref{sols}), resulting in
\begin{eqnarray}
D(p,t) \, &=& \, \frac{1} 
{\sqrt {\pi \, \left (4 \, \int^t {\cal D}_F(t') \, \exp \left [ 
\, 2\int^{t'}{\cal A}(t'')\, dt'' \right ] \, dt' \right ) \, 
\left [\exp \left (-2 \int^{t'}{\cal A}(t'')\, dt'' \right ) \right ]}} 
\nonumber \\
&\times& \, \exp \left[-\frac{\left (p-p_0\, e^{-\int^t{\cal A}(t')\, dt'}
\right )^2} 
{\left (4 \, \int^t {\cal D}_F(t') \, \exp \left [ 
\, 2\int^{t'}{\cal A}(t'')\, dt'' \right ] \, dt' \right ) \, 
\left [\exp \left (-2 \int^{t'}{\cal A}(t'')\, dt'' \right ) \right ]} 
\right ]  \, \, . \label{solmom}
\end{eqnarray}
For relativistic particles, $p=E$, (\ref{solmom}) can be written as
\begin{eqnarray}
D(E,L) &=& \frac{1}{\sqrt{\pi\, {\cal W}(L)}} \, \exp \left [ 
- \frac{\left (E-E_0\, e^{-\int^L_0{\cal A}(t') \, dt'} \right )^2}
{{\cal W}(L)} \right ] \, \, , \label{solfin} 
\end{eqnarray}
where ${\cal W}(L)$ is given  by
\begin {equation}
{\cal W}(L) = \left ({4\int_0^L 
{\cal D}_F(t')
\exp \left [ 2 \int^{t'} {\cal A}(t'')\, dt'' \right ]\, dt'}\right )
\left [{\exp \left (-2 \int_0^L {\cal A}(t')\, dt'\right )} \right ] \, \, .
\label{gauswid1}
\end{equation}
The energy loss of partons in the QGP
due to elastic collisions was estimated in Ref.~\cite{Thoma2} and we 
used the expression averaged over parton species as
\begin{equation}
-\, \frac{dE}{dL}\, = \, \frac{4}{3}\left(1+\frac{9}{4}\right )\,
\pi \alpha_s^2 T^2 \left ( 1+\frac{n_f}{6} \right ) \log \left [ 
2^{n_f/2(6+n_f)}\, 0.92 \, \frac{\sqrt{ET}}{m_g} \right ]\, , \label{pdedx}
\end{equation}
where $n_f$ is the number of quark flavours, $\alpha_s$ is the strong coupling
constant,
$m_g= \sqrt {(1+n_f/6)gT/3}$ is the thermal gluon mass, 
and $E$ is the energy of the partons.
Following (\ref{average}), we can now estimate ${\cal A}$ for different quark 
flavours and gluons at the energies (temperatures) of interest.
For averaging over the momentum the Boltzmann
distribution was used.
The time dependence of the drag coefficient comes from assuming
a temperature $T(t)$ decreasing with time as the system expands, 
according to the Bjorken scaling law~\cite{bjor} $T(t)=t_0^{1/3} T_0/t^{1/3}$, 
where $t_0$ and $T_0$ are, respectively, the initial time and temperature at 
which the background of the partonic system has attained local kinetic 
equilibrium. Since the plasma 
expands with the passage of time, 
we have used the length of the plasma, $L$ as the maximum time limit for the
relativistic case ($v\sim 1$). In Fig.~1 the drag coefficient in the 
QGP phase of the expanding fireball is shown as a function of time, where 
we have chosen the parameters $T_0=0.5$ GeV, $t_0=0.3$ fm, and $\alpha_s=0.3$.

\vspace{-0.2in}

\centerline{\psfig{figure=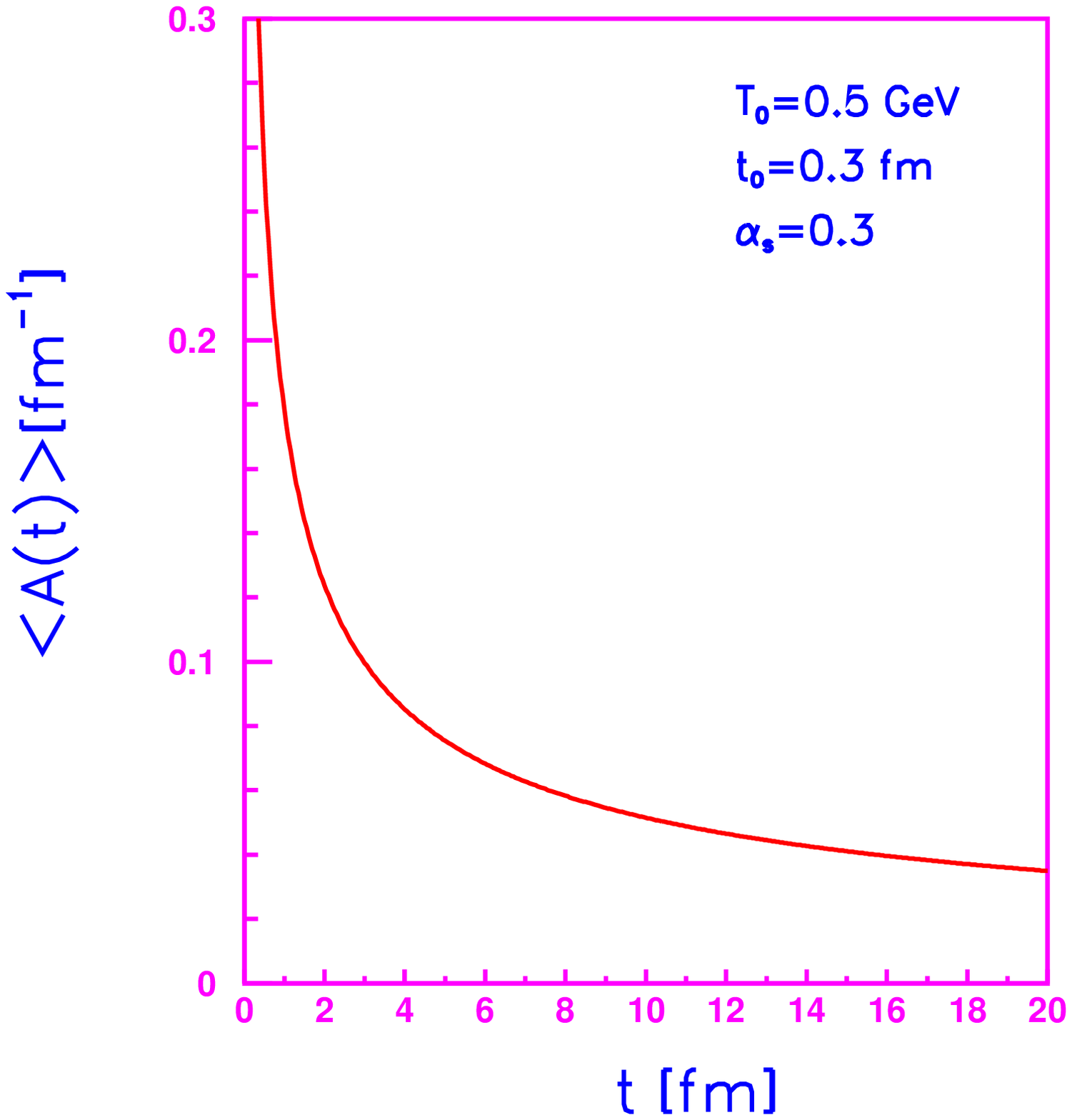,height=9cm,width=10cm}}

\vspace{-0.6in}
\noindent Figure~1: The drag coefficient ${\cal A}(t)$ in an expanding QGP.

\vspace{0.3in}

In Fig.~2 we show the probability distribution, $D(E,L)$ given in 
(\ref{solfin}), as a function of energy $E$, choosing $E_0=1$ GeV. 
The peak of the probability 
distribution is shifted with passage of time (or distance travelled) 
indicating the most probable energy loss due to elastic
collisions in the medium. 

\vspace{-0.2in}

\centerline{\psfig{figure=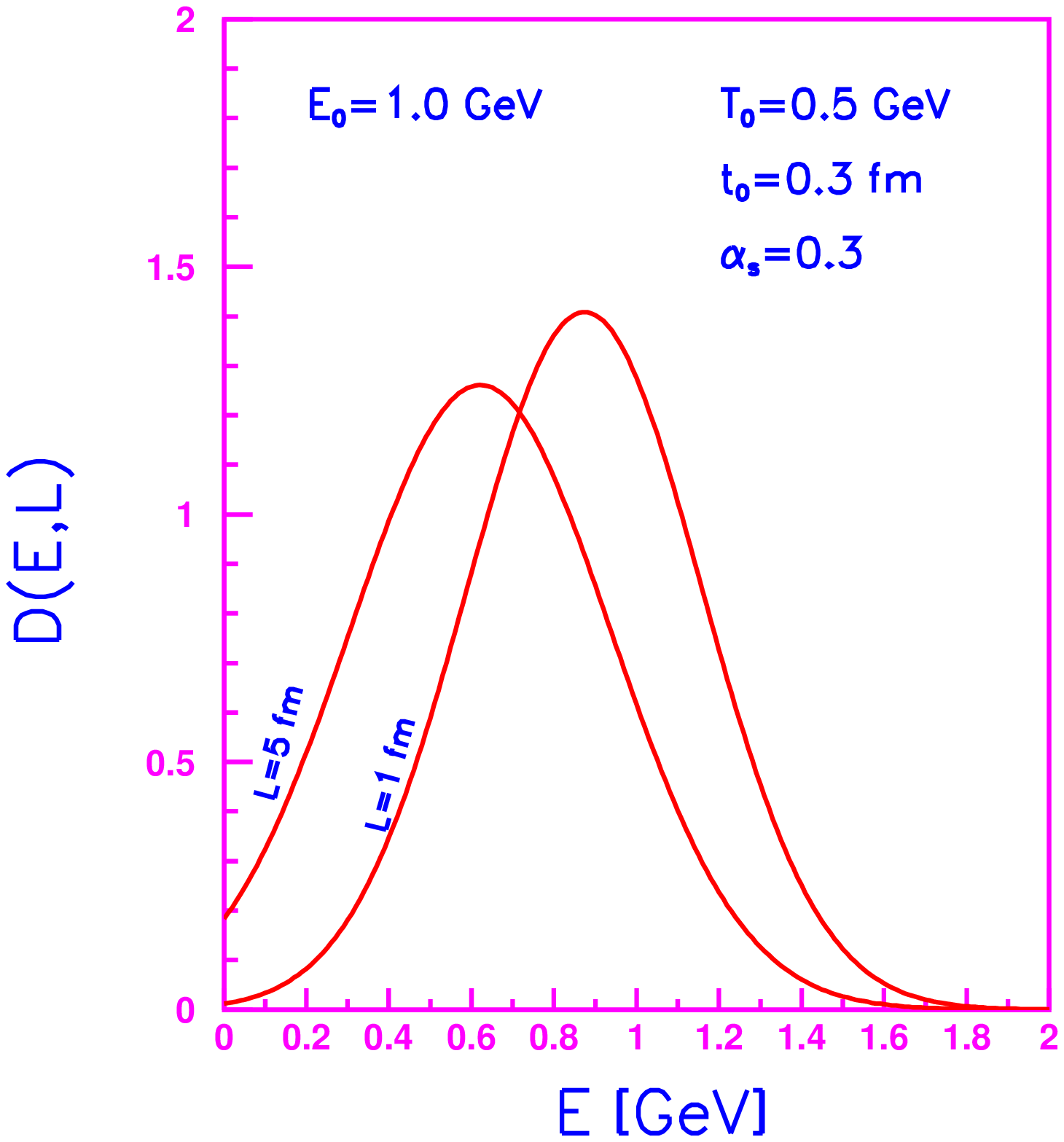,height=9cm,width=9cm}}

\vspace{-0.6in}
\noindent Figure~2: The energy-loss probability distribution, $D(E,L)$ 
as a function of energy $E$ for a given distance $L$. 

\vspace{0.3in}

After traversing a distance $L$, the mean energy of the parton due to 
the elastic collisions in the medium can be estimated as 
\begin{equation}
\langle \, E \, \rangle \, = \, \int_0^\infty \, E \, D(E,L) \, dE \, 
= \, E_0 \, e^{\, - \, \int_0^L \, {\cal A}(t') \, dt'} \,  
= \, m_\bot \, e^{\, - \, \int_0^L \, {\cal A}(t') \, dt'} \,  , 
\label{meane} 
\end{equation}
where $E_0=m_\bot \, = \sqrt{p_\bot^2 \, + m^2 }$ at the central
rapidity region, $y=0$. 
The average energy loss due to elastic collisions in the medium is 
given by
\begin{eqnarray}
\Delta E\, = \langle \, \epsilon \, \rangle \, &= &\, 
E_0\, -\, \langle \, E \, \rangle  \, \, \nonumber \\
&=& \, m_\bot \, \left ( \, 1\, -\, e^{\, - \, \int_0^L \, {\cal A}(t') 
\, dt'} \, \right ) \, \, . \label{avgelos}
\end{eqnarray}

For the massless case, the average $\Delta p_\bot$ can be written as
\begin{eqnarray}
\Delta p_\bot\, &= &\, 
 \, p_\bot \, \left ( \, 1\, -\, e^{\, - \, \int_0^L \, {\cal A}(t') 
\, dt'} \, \right ) \, \, . \label{avgmom}
\end{eqnarray}

\vspace{-0.1in}

\centerline{\psfig{figure=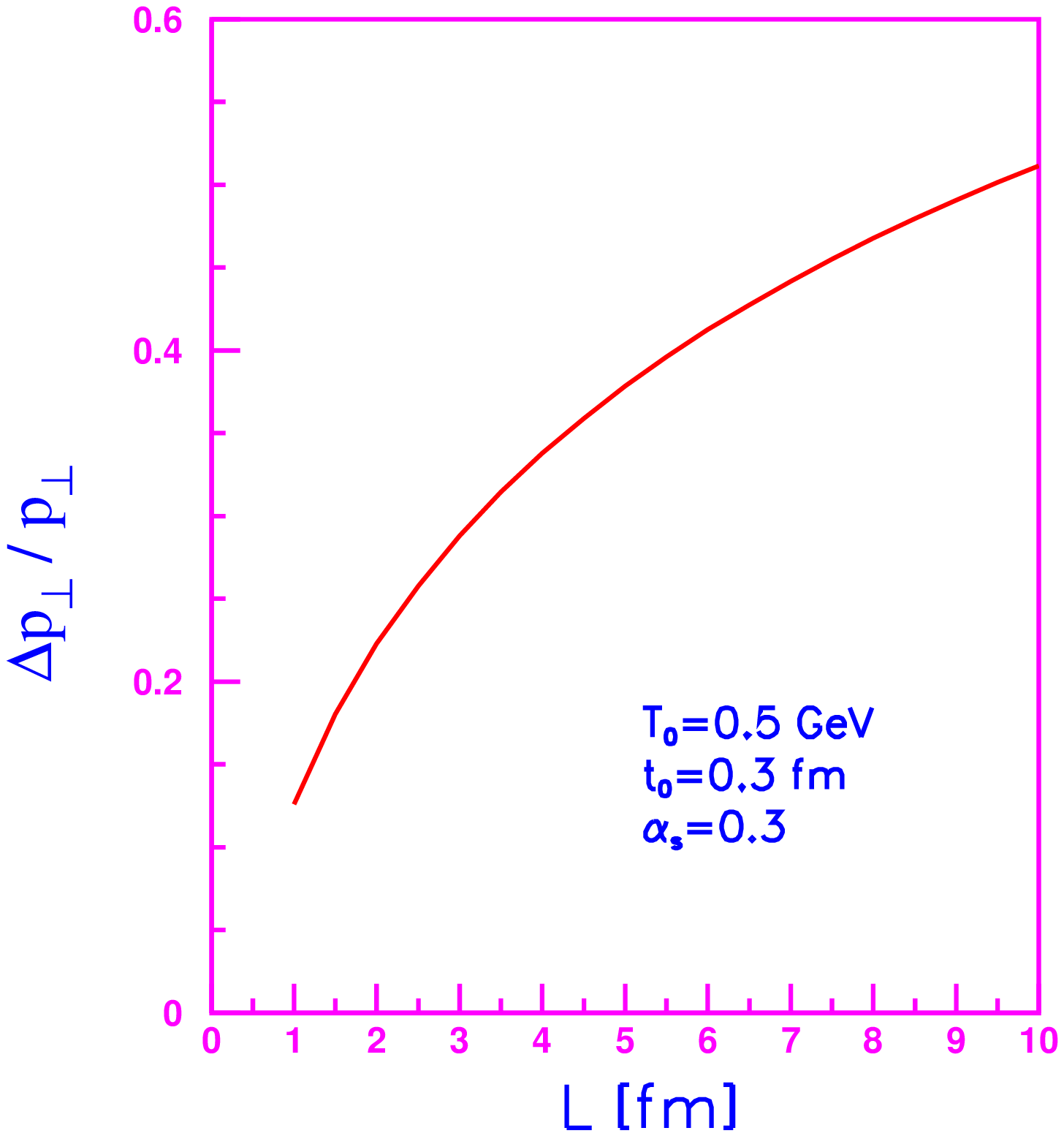,height=9cm,width=10cm}}

\vspace{-0.4in}
\noindent Figure~3: The effective shift of the scaled
transverse momentum $\Delta p_\bot/p_\bot$ as a function
of distance $L$.

\vspace{0.3in}

In Fig.~3 we display the scaled effective energy loss of fast partons due to
elastic collisions in the medium as function of the distance $L$. It is
found that the effective energy loss, defined as the shift of the momentum 
spectrum of fast partons, depends on the transverse momentum $p_\bot$ and 
the distance $L$. For given $p_\bot$ the effective energy loss is 10$\%$ after
traversing a distance of 1fm and around 50$\%$ after 10 fm. The effective 
energy-loss scales linearly with $p_\bot$ for a given $L$. 
This clearly reflects a random walk in $p_\bot$ as 
a fast parton moves in the medium~\cite{muel} with some interactions 
resulting in an energy gain and others in a loss of energy.

We assume that the geometry is described by a cylinder of radius $R$,
as in the Boost invariant Bjorken model~\cite{bjor} of nuclear collisions,
and the parton moves in the transverse plane in the local rest frame.
Then a parton created at a point $\vec{\mathbf r}$ with an angle $\phi$
in the transverse direction will travel a distance \cite{muel}

\begin{equation}
L(\phi)= (R^2\, - \, r^2 \, \sin^2\phi \,)^{1/2}\, - \, r \, \cos \phi \, \,
\, , \label{trad}
\end{equation}
where $\cos \phi\, = \,{\hat {\vec {\mathbf v}}} \, \cdot \, 
{\hat {\vec {\mathbf r}}}\, \, $; ${\vec {\mathbf v}}$ is the velocity 
of the parton and $r\, = \, |{\vec {\mathbf r}}|$.

The parameterization of the $p_\bot$ distribution~\cite{muel,dks} 
which describes the first RHIC hadroproduction data for moderately
large values of $p_\bot$ has the form
\begin{equation}
\frac{dN^{\rm{vac}}}{d^2p_\bot}\, = \, N_0 \left (1\, +\, \frac{p_\bot}{p_0}
\right )^{-\nu} \, \, \, , \label{param}
\end{equation}
where $\nu \approx 8$ and $p_0=1.75$ GeV.
The quenched  spectrum convoluted with the transverse geometry of the partonic
system can be written from (\ref{rate}) as
\begin{equation}
\frac{dN^{\rm{med}}}{d^2p_\bot}\, =
\, Q(p_\bot) \, \frac{dN^{\rm{vac}}}{d^2p_\bot}\, = \,
\frac{1}{2\pi^2 R^2} \ \int_0^{2\pi} \, d\phi \, \int_0^R \, d^2r \, \, 
\frac{dN (p_\bot+ \Delta p_\bot)}{d^2 p_\bot}\, \, \, . \label{supp}
\end{equation}

\vspace{0.3in}

\centerline{\psfig{figure=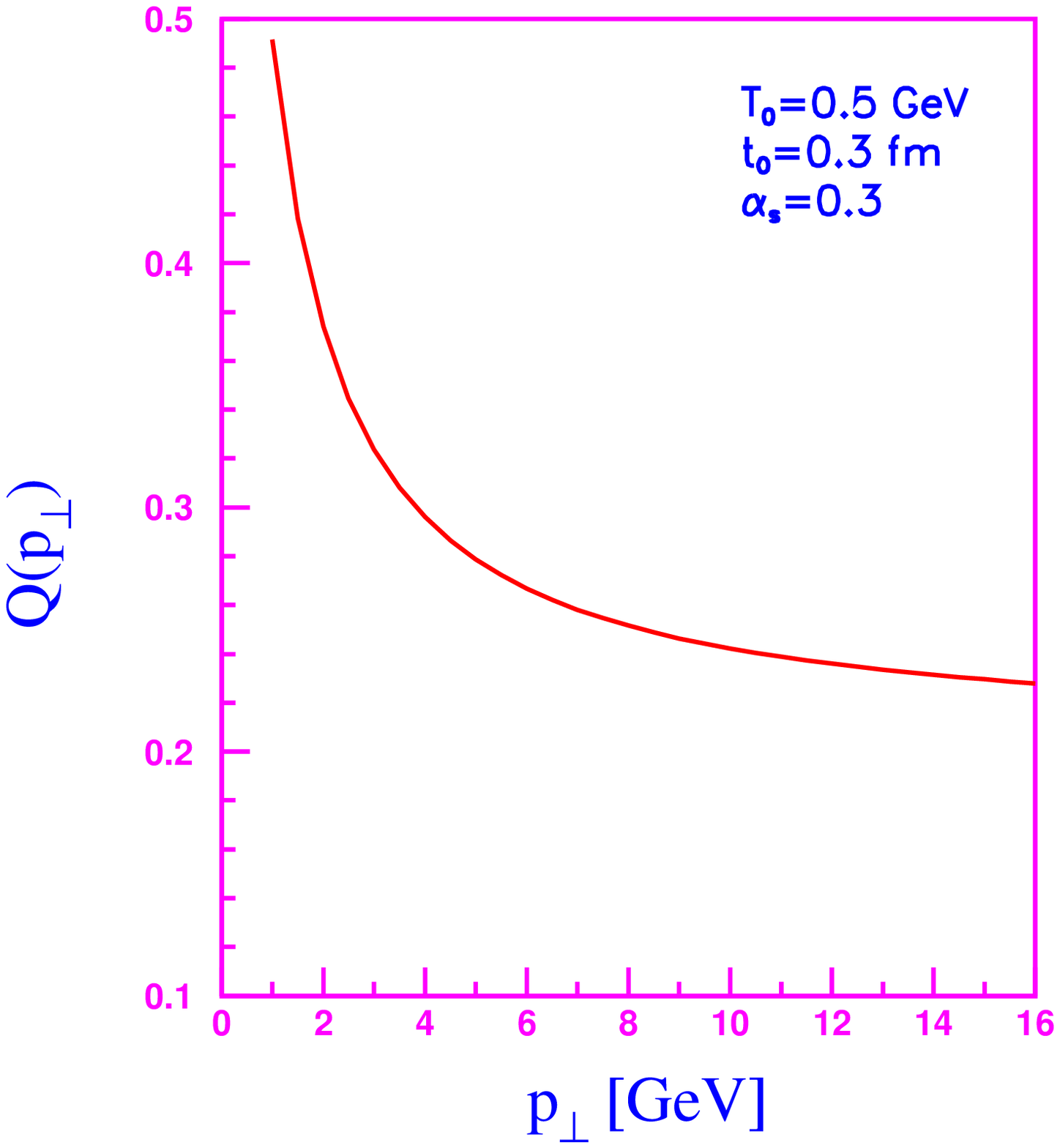,height=9cm,width=10cm}}

\vspace{-0.4in}
\noindent Figure~4: The quenching factor $Q(p_\bot)$ as a function of 
transverse momentum $p_\bot$.

\vspace{0.3in}

In Fig.~4 our numerical results of the quenching factor 
$Q(p_\bot)$ due to elastic collisions in the medium and its dependence 
on $p_\bot$ are shown. It is found that the quenching becomes stronger 
at lower $p_\bot$ and gradually decreases with higher $p_\bot$ in agreement 
with experiments \cite{RHIC}. As discussed 
above the quenching factor obtained here is quite similar to a random walk as 
well as the Bethe-Heitler scaling law given phenomenologically in 
Ref.~\cite{muel}. 

Comparing the quenching factor due to elastic collisions, shown in Fig.~3, 
with the one coming from the radiative energy loss \cite{muel} we observe 
that both are of similar magnitude. The radiative quenching factor varies 
for $p_\bot$ from 6 to 16 GeV between 0.17 and 0.24 depending on the model
assumptions, whereas the collisional quenching factor lies between 0.22 and 
0.26 for the same momentum range and for our choice of the parameters, i.e., 
$T_0=0.5$ GeV, $t_0=0.3$ fm, and $\alpha_s=0.3$. Apart from uncertainties
in these parameters let us also have a look at some of the assumptions made 
in this work which may modify the quenching factor. 
First, as discussed above, the momentum dependence of the drag coefficent, 
containing the 
dynamics of the elastic collisions, has been averaged out. A major advantage 
of this is the simplicity of the resulting differential equation. 
Of course, this simplification  
can lead to some amount of uncertainty in the quenching factor. Secondly, the 
entire discussion is based on the one dimensional Fokker-Planck equation and
the
Bjorken model of the nuclear collision, which may not be a very realistic 
description here but can provide a very intuitive picture of the 
problem.
However, extension to three dimension is indeed an ambitious goal,
which may cause that many of the considerations of the present work 
will have to be revised.
Within the limitation of our simplified model the contribution of the 
collisional energy loss of partons in the QGP to the quenching of hadron 
spectra in ultrarelativistic heavy-ion collisions is found to be
very important and cannot be neglected. In particular, if the radiative 
energy loss is suppressed at intermediate energies, the collisional 
quenching would explain the observed decreasing of the quenching factor with
increasing transverse momentum.

\vspace{0.5in}

\noindent{\bf Acknowledgment:} MGM is thankful to J. Alam, A. Dhara, 
B. M\"uller and S. Sarkar for useful discussion.

\end{document}